\documentclass[preprint2]{aastex61}

\newcommand\aastex{AAS\TeX}

\shorttitle{\aastex\ X-ray variability of Geminga's PWN}
\shortauthors{Hui et al.}

\begin{document}

\title{Rapid X-ray variations of the Geminga pulsar wind nebula}

\correspondingauthor{C. Y. Hui}
\email{cyhui@cnu.ac.kr}

\author{C. Y. Hui}
\author{Jongsu Lee}
\affil{Department of Astronomy and Space Science, Chungnam
National University, Daejeon 34134, Korea}

\author{A. K. H. Kong}
\affil{Institute of Astronomy and Department of Physics, National Tsing Hua University, Hsinchu, Taiwan}
\affil{Astrophysics, Department of Physics, University of Oxford, Keble Road, Oxford OX1 3RH, U.K.}

\author{P. H. T. Tam}
\affil{School of Physics \& Astronomy, Sun Yat-Sen University, Guangzhou 510275, China}

\author{J. Takata}
\affil{Institute of Particle physics and Astronomy, Huazhong University of Science and Technology, China}

\author{K. S. Cheng}
\affil{Department of Physics, University of Hong Kong, Pokfulam Road, Hong Kong}

\author{Dongsu Ryu}
\affil{Department of Physics, UNIST, Ulsan 44919, Korea}
\affil{Korea Astronomy and Space Science Institute, Daejeon 34055, Korea}
\begin{abstract}
A recent study by Posselt et al. (2017) reported the deepest 
X-ray investigation of the Geminga pulsar wind nebula (PWN) by using \emph{Chandra X-ray Observatory}. In comparison with previous 
studies of this system, a number of new findings have been reported and we found these suggest the possible variabilities
in various components of this PWN. This motivates us to carry out a dedicated search for the morphological and 
spectral variations of this complex nebula.
We have discovered variabilities on timescales from a few days to a few months from different components of the nebula. 
The fastest change occurred in the circumstellar environment at a rate of 80 per cent of the speed of light. 
One of the most spectacular results is the wiggling of a half light-year long tail as an 
extension of the jet, which is significantly bent by the ram pressure. The jet wiggling occurred at a rate of 
about 20 per cent of the speed of light. This twisted structure can possibly be a result of a propagating 
torsional Alf\`{v}en wave. We have also found evidence of spectral hardening along this tail for 
a period of about nine months. 
\end{abstract}


\section{Introduction} 
It is generally believed that most of the rotational energy from a pulsar are carried away by a
relativistic particle outflow which is known as pulsar wind (Gold 1969). When these fast-moving wind 
particles interact with the interstellar medium, shocked emission can be formed leading to the formation of 
pulsar wind nebulae (PWNe) (Rees \& Gunn 1974; Arons 2012). Theoretical models suggest that the morphology 
of PWNe is manipulated by magneto-hydrodynamic (MHD) instability of the outflows (Mizuno et al. 2014; 
Singh et al. 2016). Therefore, studying the variabilities of PWNe provides a means to investigate the underlying 
MHD activities. 

Geminga (PSR J0633+1746) is a nearby ($d\sim250$~pc) middle-age ($\tau=3.4\times10^{5}$~yrs) 
radio-quiet $\gamma-$ray pulsar with a spin-down power of $\dot{E}=3.3\times10^{34}$~erg/s 
(Faherty et al. 2007; Manchester et al. 2005). Its projected velocity 
of $v\sim211$~km/s suggests that Geminga is in a supersonic motion and it should be associated with 
a bow shock PWN. Given its proximity, it is expected to provide an ideal system for resolving and investigating 
various components of the shock emission.

With the first \emph{Chandra} observation of Geminga in 2004,
Sanwal et al. (2004) and Pavlov et al. (2006) have found an axial tail that extends at least up to
$\sim25^{"}$ directly behind Geminga's proper motion direction.
In revisiting this system with the second \emph{Chandra} observation in 2007, Pavlov et al. (2010) 
have shown that the axial tail extends for $\sim50^{"}$ from the pulsar. Several blobs have also been 
found on the axial tail. Comparing the images obtained in 2004 and 2007, Pavlov et al. (2010) have 
noticed the possible variability of Geminga's axial tail in terms of its spectrum, brightness 
and the appearance of blobs along it.
 
With this deeper observation, two faint $\sim2^{'}$ long outer tails as firstly found by Caraveo et al. (2003) 
with \emph{XMM-Newton} are also detected. Both tails are significantly bent away from the direction of proper motion.

Very recently, Posselt et al. (2017) have reported a detailed study of Geminga's PWN with twelve more new 
\emph{Chandra} observations carried out from 2012 November to 2013 September. 
Based on the spatial analysis, they have confirmed the morphological variability of the 
axial tail as reported by Pavlov et al. (2010). However, only the combined spectrum has been examined in their study.
It remains unknown whether there is any associated spectral variability. 

Also, Posselt et al. (2017) has found the combined spectrum of southern outer tail is significantly harder than that of the 
axial tail (cf. Tab.~3 in Posselt et al. 2017). This is different from the results reported by Pavlov et al. (2010) 
which found the photon indices of the southern tail and the axial tail are comparable in the earlier observations in 
2004 and 2007 (cf. Tab.~1 in Pavlov et al. 2010). This indicates the possible variability in the southern tail such 
that its spectrum was hardened in certain epoch.

Another interesting feature reported by Posselt et al. (2017) is the new component found in the immediate surrounding of 
the pulsar (cf. Fig.~6 in their paper). The most prominent one is the southeastern feature that connects the pulsar and 
the southern outer tail. This component can only be found in several epochs. This also suggests the possible variability of 
this component. However, the significance of the feature is not mentioned in Posselt et al. (2017). It remains unclear 
if this small-scale feature is genuine or merely a result of background fluctuation. 

Motivated by all these new findings, we have performed a follow-up investigation of this complex X-ray nebula with an aim to 
search for the variabilities in different components.

\section{OBSERVATION \& DATA REDUCTION}

All the data used in this investigation were obtained by the Advanced CCD Imaging Spectrometer (ACIS)
on board \emph{Chandra}. The details of these observations are summarized in Table~1.
For the data reduction,
we utilized the Chandra Interactive Analysis of Observations software (CIAO) with the updated calibration
data base (CALDB). Using the CIAO script \emph{chandra$\_$repro}, we reprocessed all the data with
subpixel event repositioning in order to facilitate a high angular resolution analysis. 
Since the instrumental background increases sharply below 0.5 keV on the front-illuminated CCDs (ACIS-I) which were 
used in all observations except the one in 2004, we apply the low energy cut at 0.5~keV.  
All the investigations in this work were restricted in an energy range of $0.5-7$~keV.

To begin with, we have performed a preliminary screening of all the full-field images from individual observations
by visual inspection.
For those observations separated by less than 10 days, we do not find any obvious variability among their
images in this preliminary check. In view of this, we group the data into 8 different epochs as defined in
Table~1 so as to optimize the photon statistic for the subsequent analysis. Hereafter, we adopt the definition
of these epochs for referring the the time frame in our analysis.
With this scheme of data grouping, the effective exposures from Epoch 3 to Epoch 8 are found to be comparable ($\sim100$~ks).
This scheme is consistent with that adopted by Posselt et al. (2017). 

In this investigation, we only consider any spatial features that detected at a signal-to-noise ($S/N$) ratio $>5\sigma$ as
genuine. This criterion allows us to better discriminate the variable diffuse emission from the background fluctuations.

We followed the method in Li \& Ma (1983) in calculating the S/N ratios for all features in our work. The net source counts $N_{\rm net}$ in a 
region-of-interest of an area $A_{S}$ is estimated by $N_{\rm net}=N_{\rm tot}-\alpha N_{\rm bkg}$, where $N_{\rm tot}$ is the total counts 
sampled in $A_{S}$, $N_{\rm bkg}$ is the background counts sampled 
in a source-free region of an area $A_{B}$ and $\alpha=A_{S}/A_{B}$. The estimate of the standard deviation of $N_{\rm net}$ is given by 

\begin{equation}
\hat{\sigma}\left(N_{\rm net}\right)=\sqrt{N_{\rm net}+\left(1+\alpha\right)\alpha N_{\rm bkg}}. 
\end{equation}

And hence, $S/N$ is estimated as $N_{\rm net}/\hat{\sigma}\left(N_{\rm net}\right)$.

\begin{table}
\caption{Details of \emph{Chandra} observations of the field of Geminga in different epochs.}
\begin{center}
\resizebox{!}{5cm}{
\begin{tabular}{l c c c c}
\hline
\hline
Obs. ID. & Detector & Start Date \& Time & Mode &  Exposure \\
         &          &  (UTC)             &      &      (ks)           \\
\hline
\multicolumn{5}{c}{\bf Epoch 1} \\
\hline
4674    & ACIS-S   &  2004-02-07 13:01:59  & FAINT & 18.8 \\
\hline
\multicolumn{5}{c}{\bf Epoch 2} \\
\hline
7592   & ACIS-I   &  2007-08-27 10:10:05  & VFAINT & 77.1 \\
\hline
\multicolumn{5}{c}{\bf Epoch 3} \\
\hline
15595   & ACIS-I   &  2012-11-28 22:18:14  & VFAINT & 62.24 \\
14691   & ACIS-I   &  2012-12-01 17:09:00  & VFAINT & 36.58 \\
\hline
\multicolumn{5}{c}{\bf Epoch 4} \\
\hline
14692   & ACIS-I   &  2013-01-25 07:40:57  & VFAINT & 103.68 \\
\hline
\multicolumn{5}{c}{\bf Epoch 5} \\
\hline
15623   & ACIS-I   &  2013-03-19 05:42:00   & VFAINT & 23.75 \\
15622   & ACIS-I   &  2013-03-24 12:10:42   & VFAINT & 47.04 \\
14693   & ACIS-I   &  2013-03-27 03:49:08   & VFAINT & 22.76 \\
\hline
\multicolumn{5}{c}{\bf Epoch 6} \\
\hline
14694   & ACIS-I   &  2013-04-22 17:37:27 & VFAINT & 96.25 \\
\hline
\multicolumn{5}{c}{\bf Epoch 7} \\
\hline
15551   & ACIS-I   & 2013-08-25 20:12:33  & VFAINT & 30.66 \\
16318   & ACIS-I   & 2013-08-28 13:21:30  & VFAINT & 19.81 \\
16319   & ACIS-I   & 2013-08-30 12:31:46  & VFAINT & 44.48 \\
\hline
\multicolumn{5}{c}{\bf Epoch 8} \\
\hline
15552   & ACIS-I   & 2013-09-16 03:42:22  & VFAINT & 37.08 \\
16372   & ACIS-I   & 2013-09-20 13:17:03  & VFAINT & 59.28 \\
\hline
\end{tabular}
}
\end{center}
\end{table}

\section{VARIABILITY ANALYSIS OF VARIOUS SPATIAL COMPONENTS}
\begin{figure*}
\plotone{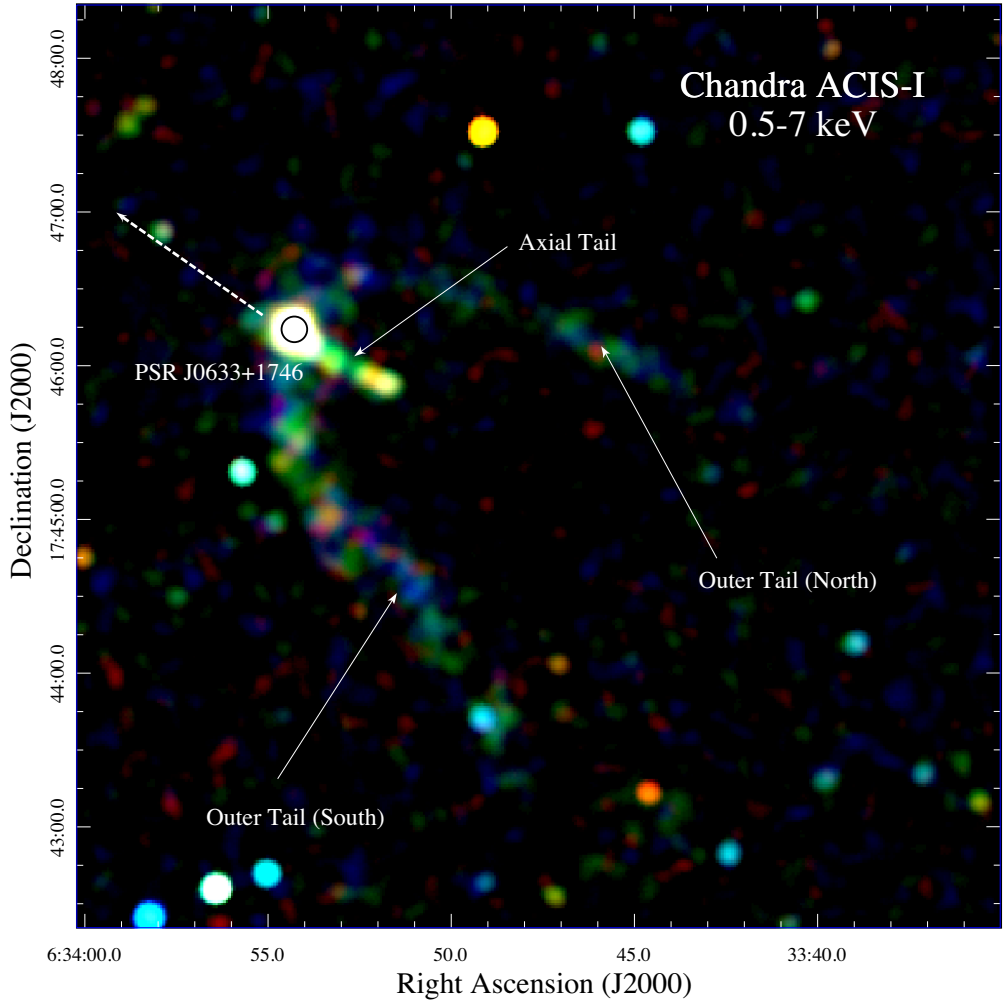}
\caption{A $6^{'}\times6^{'}$ deep X-ray colour image that combines all \emph{Chandra} data except for the one
obtained in 2004 (red: 0.5-1 keV, green 1-2 keV, blue 2-7 keV). The effective exposure of this image is $\sim660$~ks. The image is smoothed with a Gaussian of $\sigma=6^{"}$.
Various components of the pulsar wind nebula (i.e. axial tail and the outer tails) are illustrated accordingly.
The dashed arrow shows the direction of Geminga's proper motion.}
\label{feature}
\end{figure*}

Figure~\ref{feature} shows the $6^{'}\times6^{'}$ field of Geminga and its PWN, which combines all the X-ray observations that
cover the entire PWN. Apart from the pulsar as the brightest point-like source, there are two distinct extended structures near the pulsar --- 
axial tail and outer tails, as illustrated in Figure~\ref{feature}.
Axial tail is intimately connected to the pulsar which spans an apparent length of $\sim42^{"}$ (0.05 pc) from the pulsar position 
to a bright blob at the end of this feature (i.e. Blob C as illustrated in Fig.~\ref{axial} and Fig.~\ref{profile}) in this combined image.
Besides this bright blob, the axial tail also appears to comprise several clumpy structures (Please refer to \S3.1 for further 
details). 
Besides the axial tail, two outer tails extending out to about $2^{'}$ (0.15 pc) can be clearly seen. A detailed analysis of these two long tails will be 
presented in \S3.3.

From this energy-coded colour image, we notice that the X-rays from both outer tails are harder than those from the axial tail. 
This is consistent with the results reported by Posselt et al. (2017) based on their analysis of combined spectra. 
For the southern outer tail, we also note that the X-rays are apparently harder in the region further away from the pulsar. 

In the followings, we investigate the variability of each component through a detailed spectral imaging analysis.

\subsection{Axial Tail}
\begin{figure}
\plotone{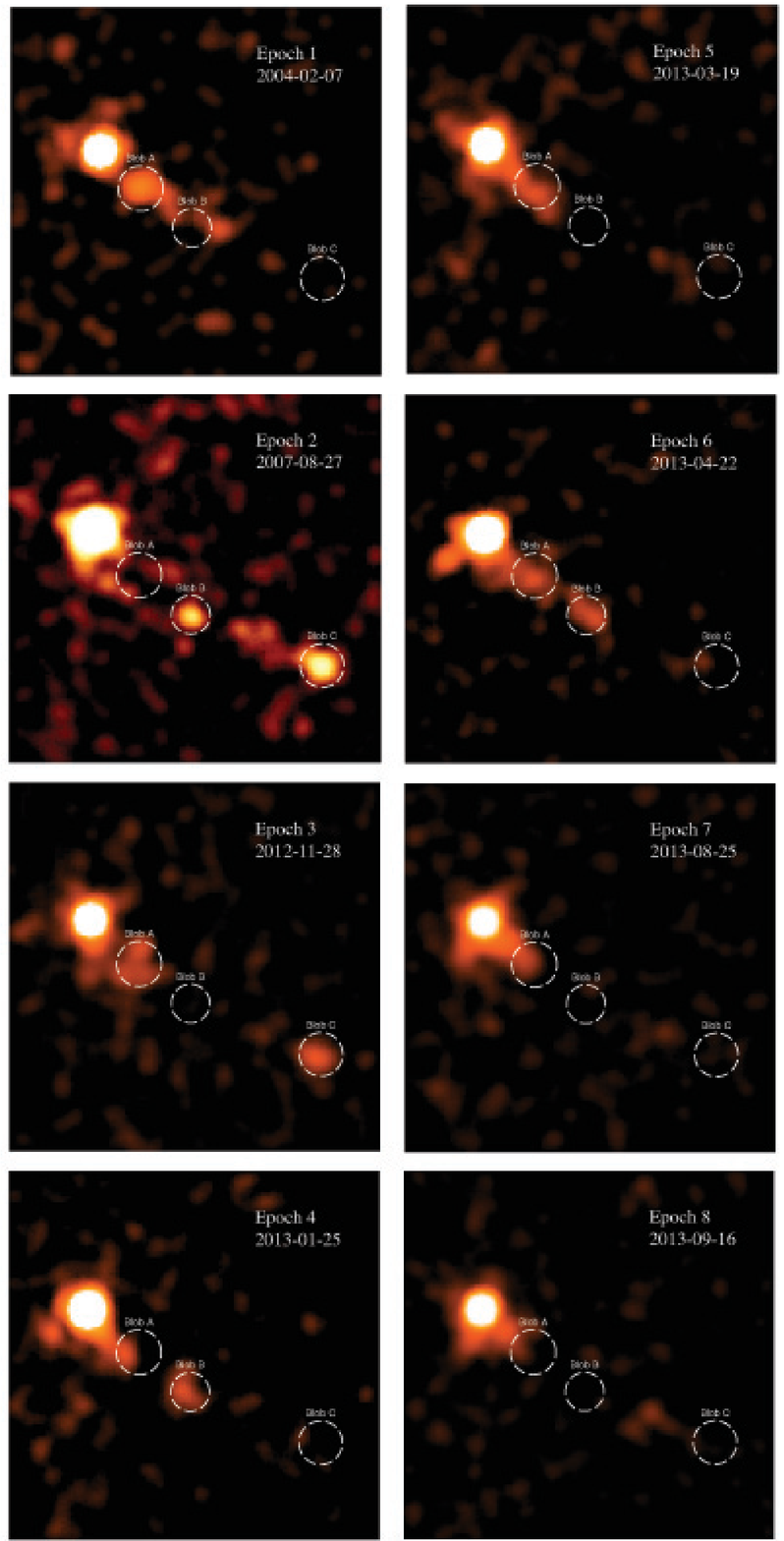}
\caption{A time sequence of \emph{Chandra} $1^{'}\times1^{'}$ exposure-corrected
images (0.5-7~keV) of the field around Geminga's axial tail ($0.5^{"}\times0.5^{"}$ for one pixel).
The start date of the corresponding epoch is given in each image.
All images are smoothed wtih a Gaussian kernel of $\sigma=2.5^{"}$.
The dashed circles illustrate the locations of Blobs A, B and C detected previously (Pavlov et al. 2010).
See also Movie 1 in the online supplementary material for the animated sequence.}
\label{axial}
\end{figure}

In this work, we define the axial tail as a feature that intimately connected to the pulsar.
Figure~\ref{axial} clearly shows the variability of the axial tail. All the images in this work are exposure-corrected
by using the CIAO script {\it fluximage}. We also overlaid the locations of Blobs A, B and C
as found by Pavlov et al. (2010) on all images for comparison.

\begin{figure*}
\plotone{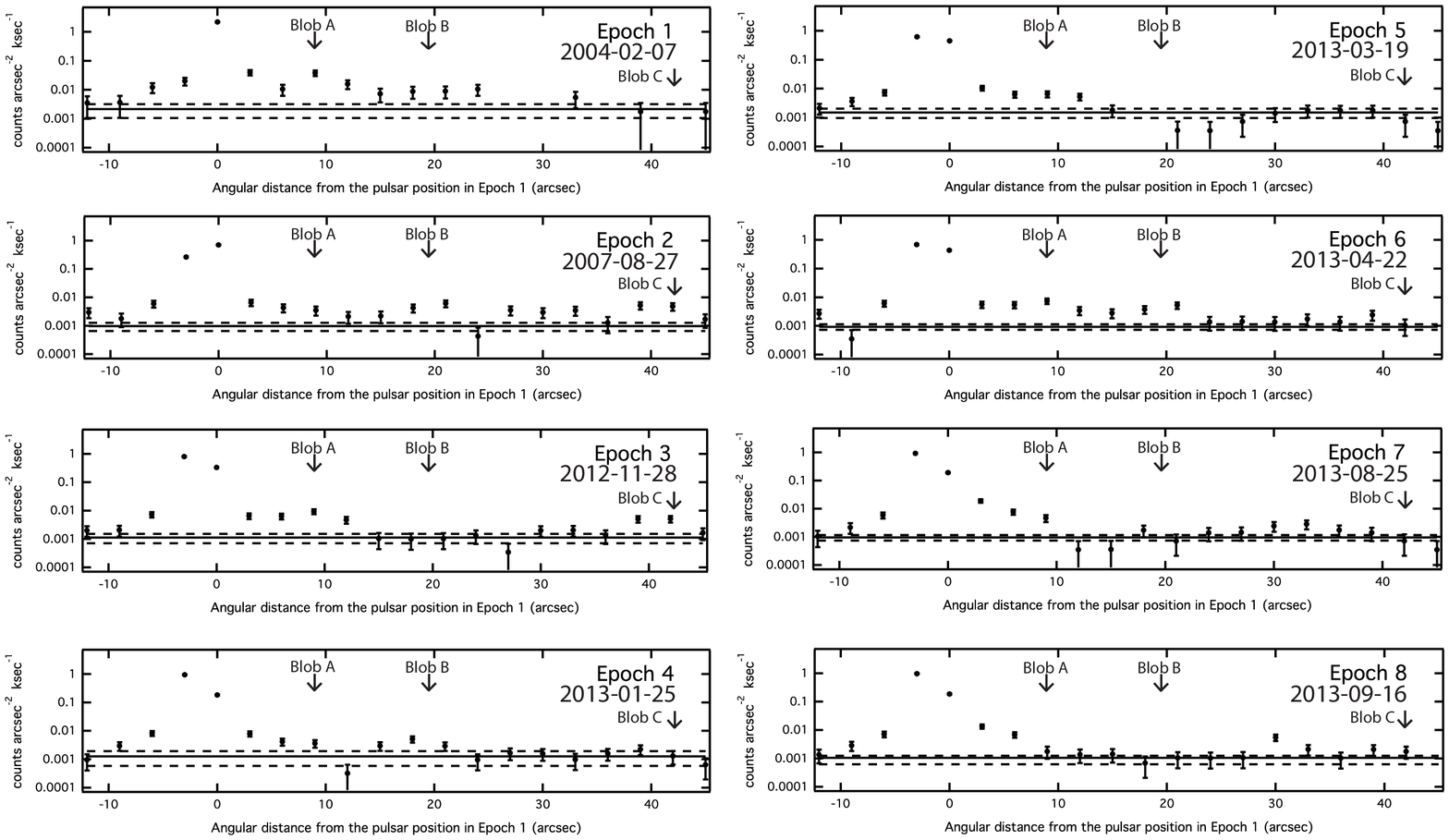}
\caption{The brightness profiles of the axial tail as computed from a $60^{"}\times10^{"}$ along the feature from
the unsmoothed images in different epochs.
The opposed direction of the pulsar's proper motion is defined as positive in the $x$ axis.
The horizontal solid and dashed lines represent the background and its $1\sigma$ uncertainties at each epoch. 
The locations of Blobs A, B and C as shown in Fig.~\ref{axial} are illustrated by the arrows accordingly.}
\label{profile}
\end{figure*}

For quantifying their lengths in each epoch, we computed their
brightness profiles by using the unsmoothed image with bin size of $0.5^{"}\times0.5^{"}$ in a $60^{"}\times10^{"}$ region along
the tail, which is large enough to cover all the identified features for the axial in different epochs.
The results are shown in Figure~\ref{profile}. The peak of the profiles corresponds to the location of the pulsar. The
shift of the peak among these profiles is a result of the proper motion of Geminga.
For the background level in each epoch, we randomly sampled in ten source-free regions of
the corresponding image and computed the medium and the $1\sigma$ uncertainties
which are illustrated by the solid and dashed horizontal lines in Figure~\ref{profile} respectively.
The length of the axial tail in each epoch is estimated by the profile of the feature extended behind the pulsar before it falls into the background.

The variability of the axial tail and the blobs are obvious from Figure~\ref{axial} and Figure~\ref{profile}. Since the 
morphological changes of these feature have been studied in details by Posselt et al. (2017), we do not repeat this analysis. 
And we would like to refer the interested readers to Sec.~2.4 in their paper. 

In our work, we focus in examining if there is temporal variation of its spectral properties. 
We extracted the source spectrum of the axial tail from each observation according to its extent defined by the brightness profile (Fig.~\ref{profile}) 
in the corresponding epoch. For the sake of consistency, we followed the definition the tail is a feature that intimately connected to the
pulsar. We extracted the spectrum using a single box that covers the feature and excluded those features that apparently
disconnected from it. For example, Blob B is included in the data of Epochs 2 and 6 but it is excluded in Epoch 4.
The background spectra were sampled from the source free regions in the individual data.
We utilized the CIAO tool \emph{specextract} to extract all the spectra and to compute the response files.

For the spectral analysis, we used XSPEC throughout this investigation. For obtaining the results that
can be compared to those reported by Pavlov et al. (2010) and Posselt et al. (2017), we followed their procedures in the spectral fitting. We fitted
all the spectra with an absorbed power-law model with the column absorption fixed at $N_{H}=1.1\times10^{20}$~cm$^{-2}$
(Halpern \& Wang 1997; De Luca et al. 2005). All the observed fluxes and the absorption-corrected luminosities quoted in
this work are computed in 0.3-8~keV to make the comparison with the results of Pavlov et al. (2010) and Posselt et al. (2017) 
easier.

\begin{figure*}
\plotone{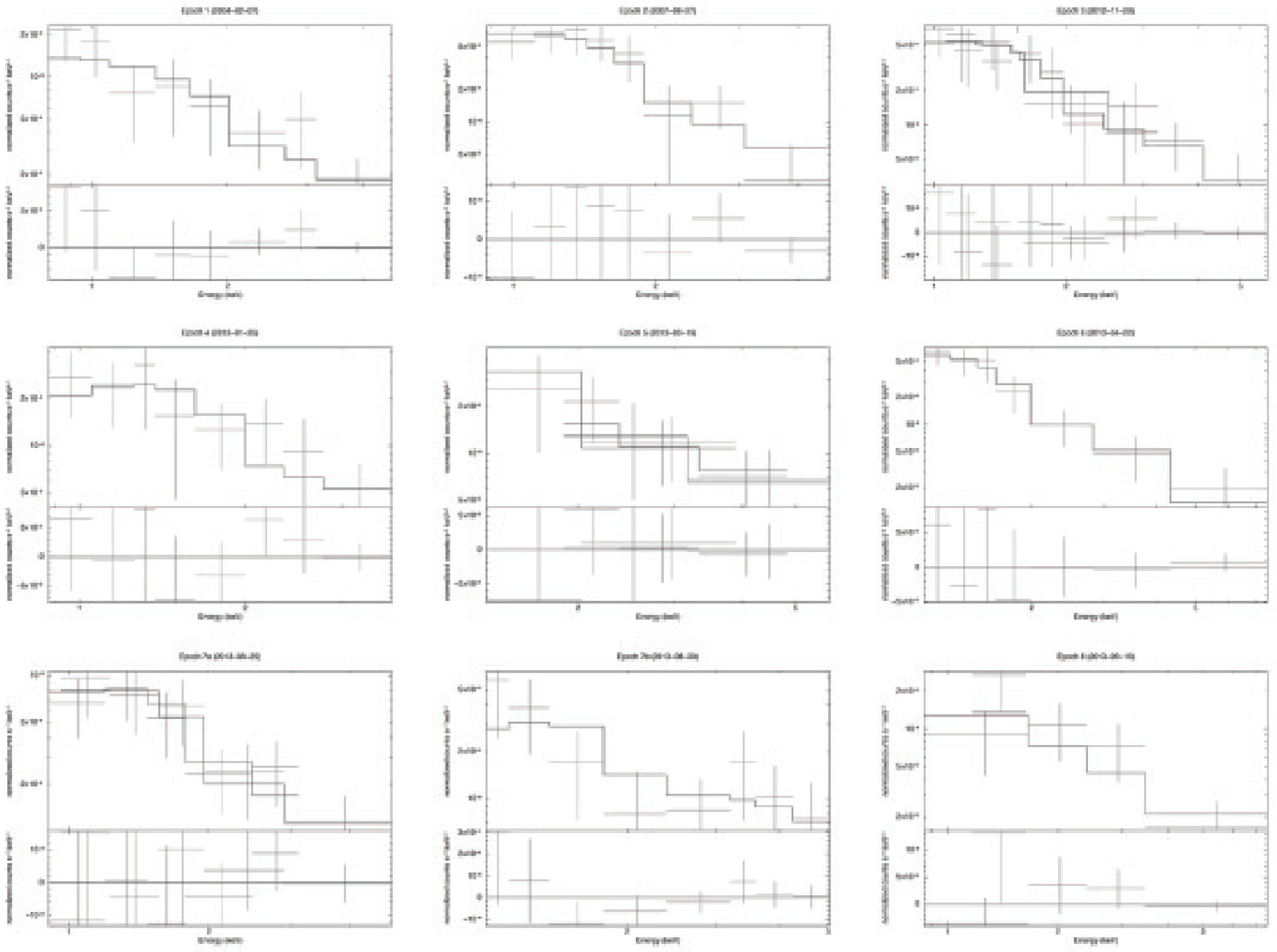}
\caption{Background-subtracted energy spectra of the axial tail ({\it upper panels}) with the best-fit power-law models (solid line) in different epochs. 
The best-fit spectral parameters are given in Tab.~\ref{axial_spec}. 
The fitting residuals are also shown ({\it lower panels}).}
\label{axial_plot}
\end{figure*}

\begin{figure}
\plotone{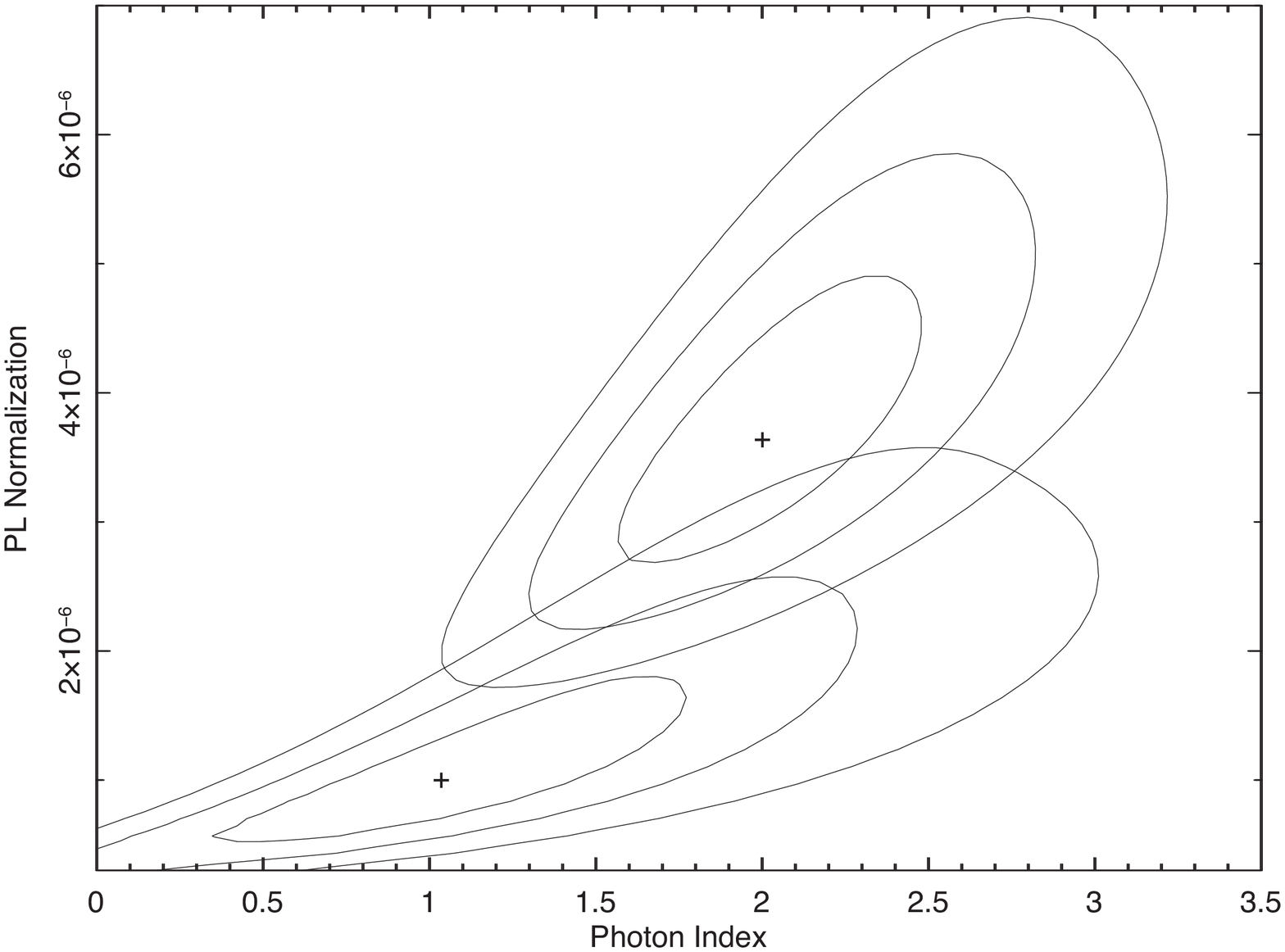}
\caption{$1\sigma$, $2\sigma$ and $3\sigma$ confidence contours for the absorbed power-law model fitted to the X-ray spectra
of Gemimga's axial tail as observed by {\it Chandra} in Epoch 7a ({\it upper contours}) and Epoch 7b
({\it lower contours}). The unit of the model normalization is photon keV$^{-1}$ cm$^{-2}$ s$^{-1}$ at 1 keV.}
\label{contour}
\end{figure}

The spectral properties of the axial tail in different epochs are summarized in Table~\ref{axial_spec}. The background-subtracted spectra and the 
fitting residuals are shown in Figure~\ref{axial_plot}.
Our results in Epochs 1 and 2 agree with those reported by Pavlov et al. (2010). Comparing the photon
indices inferred from various epochs, there is no significant variation within the tolerance of their $1\sigma$ uncertainties.
In Epoch 7, we found that the spectral results obtained individually from the data obtain in 2013 August 25 (Obs.ID: 15551) and
2013 August 28 (Obs.ID: 16318) are consistent. However, such results are not consistent with those obtained
in 2013 August 30 (Obs.ID: 16319). We have re-examined all spectra from Epoch 7 with backgrounds selected from a set of different nearby 
source-free regions. And we found the difference persists. 
It is not possible to properly characterize the spectral properties of the axial tail in this epoch by combining all data.
This leads us to further divide Epoch 7 into two sub-frames, namely Epoch 7a (Obs.ID: 15551 + Obs.ID: 16318)
and Epoch 7b (Obs.ID: 16319). We compare the confidence contours in the parameter space spanned by the photon index and the model normalization
for these two sub-frames in Figure~\ref{contour}.
It is interesting to note that the X-ray spectrum of axial tail can vary on a timescale as short as a few days.

\begin{table*}
\caption{Spectral properties of the axial tail in different epochs.}
\begin{center}
\begin{tabular}{l c c c c c c c c c}
\hline
\hline
  &  $A_{S}$  & $N_{\rm net}$  & $N_{\rm bkg}$ & $\alpha$ & S/N & $\Gamma$ & $F_{\rm obs}^{\rm a}$ & $L_{\rm unabsorb}^{\rm a}$ \\
  &   arcsec$^{2}$     &   counts   & counts &   &  $\sigma$   &          & $10^{-14}$~erg~cm$^{-2}$~s$^{-1}$ & $10^{29}$~erg/s\\
\hline
Epoch 1 & 120.8  & 44 & 7 & 0.77 & 6.0 & $1.27^{+0.46}_{-0.40}$ & $2.40^{+0.74}_{-0.61}$  & $1.82^{+0.55}_{-0.45}$ \\
Epoch 2 & 337.9  & 65 & 42 & 0.93 & 5.5  & $1.98^{+0.40}_{-0.34}$ & $1.12^{+0.20}_{-0.19}$ & $0.91^{+0.15}_{-0.14}$ \\
Epoch 3 & 108.9  & 80 & 14 & 1.00 & 7.7 & $1.92^{+0.43}_{-0.36}$ & $1.08\pm0.15$ & $0.84\pm0.12$ \\
Epoch 4 & 62.7   & 42 & 9 & 1.01 & 5.4 & $1.46^{+0.54}_{-0.50}$ & $0.64^{+0.21}_{-0.15}$ & $0.49^{+0.16}_{-0.11}$ \\
Epoch 5 & 102.1  & 63 & 9 & 1.00 & 7.0 & $1.41^{+0.57}_{-0.61}$ & $1.19^{+0.24}_{-0.23}$ & $0.90^{+0.20}_{-0.18}$ \\
Epoch 6 & 164.8  & 69 & 16 & 1.00 & 6.9 & $2.21^{+0.48}_{-0.39}$ & $1.10^{+0.26}_{-0.19}$ & $0.87^{+0.23}_{-0.15}$ \\
Epoch 7a & 60.8  & 68 & 9 & 0.47  & 7.9 & $1.69^{+0.42}_{-0.37}$ & $1.90^{+0.37}_{-0.30}$ & $1.47^{+0.27}_{-0.22}$ \\
Epoch 7b & 60.5  & 29 & 4 & 0.44 & 5.2 & $1.04^{+0.48}_{-0.46}$ &  $1.15^{+0.40}_{-0.28}$ & $0.87^{+0.29}_{-0.21}$ \\
Epoch 8 & 38.0 3 & 45 & 10 & 0.33 & 6.4 & $1.09^{+0.44}_{-0.40}$ & $0.50^{+0.15}_{-0.14}$ & $0.38\pm0.11$ \\
\hline
\end{tabular}
\end{center}
\footnotesize{$(a)$ The observed fluxes $F_{\rm obs}$ and the absorption-corrected luminosities
$L_{\rm unabsorb}$ (adopting $d=250$~pc) are computed for the energy range of 0.3-8~keV.}
\label{axial_spec}
\end{table*}

However, we noted that the $2\sigma$ and $3\sigma$ contours in these two sub-frames are overlapped.
For further investigating the variability, we examined the individual images in Epochs 7a and 7b separately to see if there is any 
corresponding morphological variation.
We examined the images of these two frames in a $30^{"}\times30^{"}$ region centered on the pulsar. To enable a more sensitive
search for the morphological variation of a small-scale feature, we produced the images with sub-pixel resolution with a
binning factor of 0.5. 
By comparing these two exposure-corrected images which are both adjusted to the same logarithmic color scale
(upper panels of Figure~\ref{subpix}), the tail behind
proper-motion direction appears to dim in less than five days. Such result is consistent with that based on the spectral analysis
and support the presence of such short timescale variability.

In view of the discovery of fast variability in Epoch 7, we have also examined the possible spectral and morphological variabilities of the axial tail
among the individual observations in Epochs 3, 5 and 8. However, there is no short-term variability has been found within these epochs.

We also examined the epoch-averaged spectral properties of the axial tail so as to cross-check the results of Posselt et al. (2017). 
Instead of fitting the combined spectrum as have been done by Posselt et al. (2017), we adopted a more rigorous approach by 
fitting all twelve individual axial tail spectra simultaneously. We found the epoch-averaged spectrum can be described by a power-law with a 
photon index of $\Gamma=1.76\pm0.17$ and a normalization factor of $(1.6\pm0.2)\times10^{-6}$~photons~keV$^{-1}$~cm$^{-2}$~s$^{-1}$ at 1 keV. 
Within the tolerance of the statistical uncertainties, the results are consistent with those reported by Posselt et al. (2017) (cf. Tab.~3 in 
their paper). 

\begin{figure}
\plotone{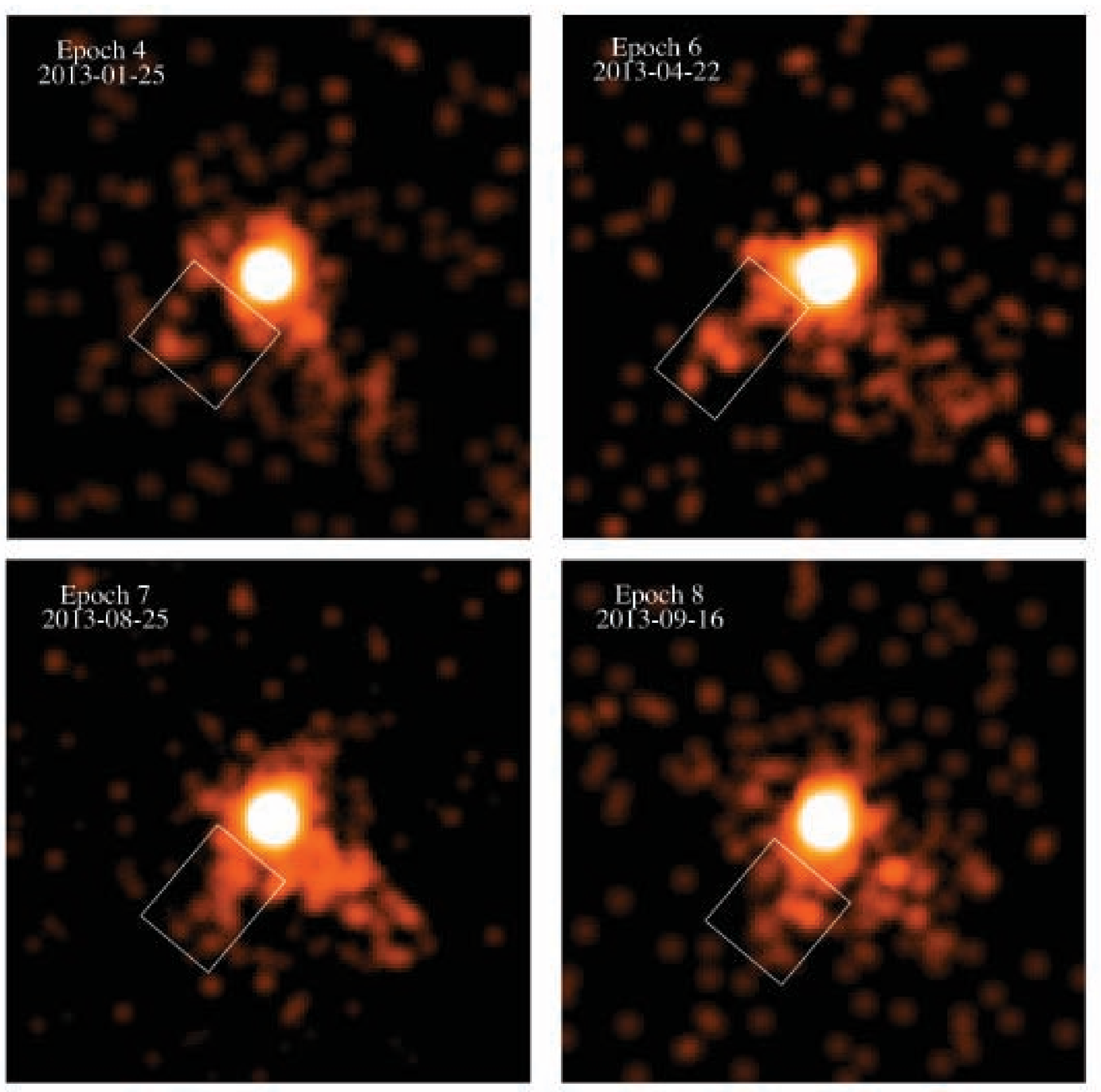}
\caption{Candidates of compact extended feature around Geminga as identified by visual inspection. The possible protrusions are 
highlighted by the boxes in each image.}
\label{jet_candidate}
\end{figure}

\begin{figure*}
\plotone{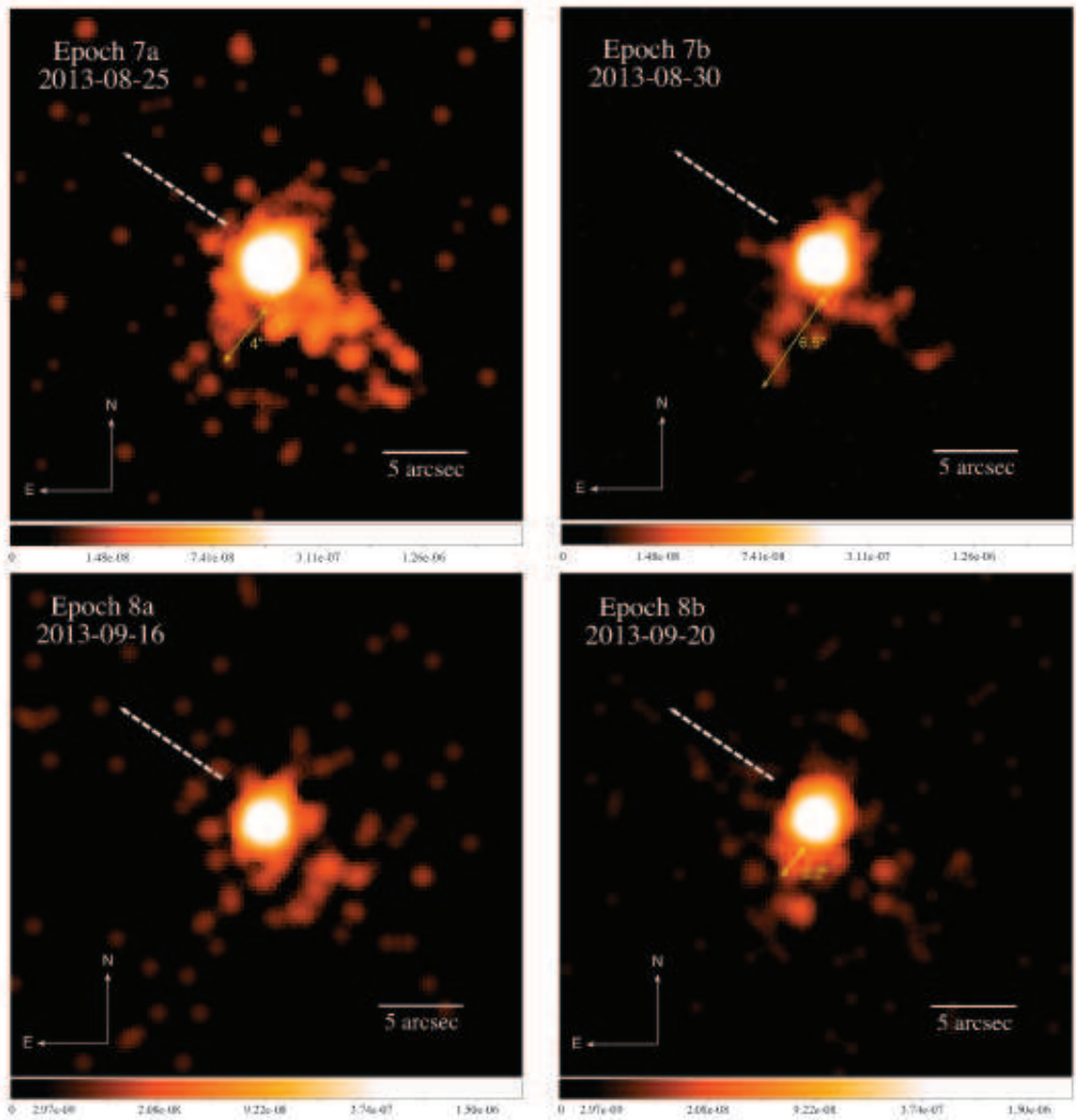}
\caption{Exposure-corrected sub-pixel resolution images ($0.25^{"}\times0.25^{"}$ for one pixel) in 0.5-7 keV of $30^{"}\times30^{"}$ centered on Geminga
as observed by {\it Chandra} in 2013 August and September. The images are smoothed by a Gaussian kernel of $1^{"}$.
The images of sub-frames in each epoch are adjusted to the same logarithmic color scale for comparison.
The morphological variation of the axial tail and the circumstellar diffuse emission can be clearly seen.
The angular sizes of the southeastern protrusion are illustrated in each images by the yellow double-arrow.
The scale bar at the bottom of each image shows the pixel values in the unit of photon cm$^{-2}$ s$^{-1}$.
See also Movie 2 and Movie 3 for the animated comparison in Epoch 7 and Epoch 8 respectively.}
\label{subpix}
\end{figure*}

\begin{figure*}
\plotone{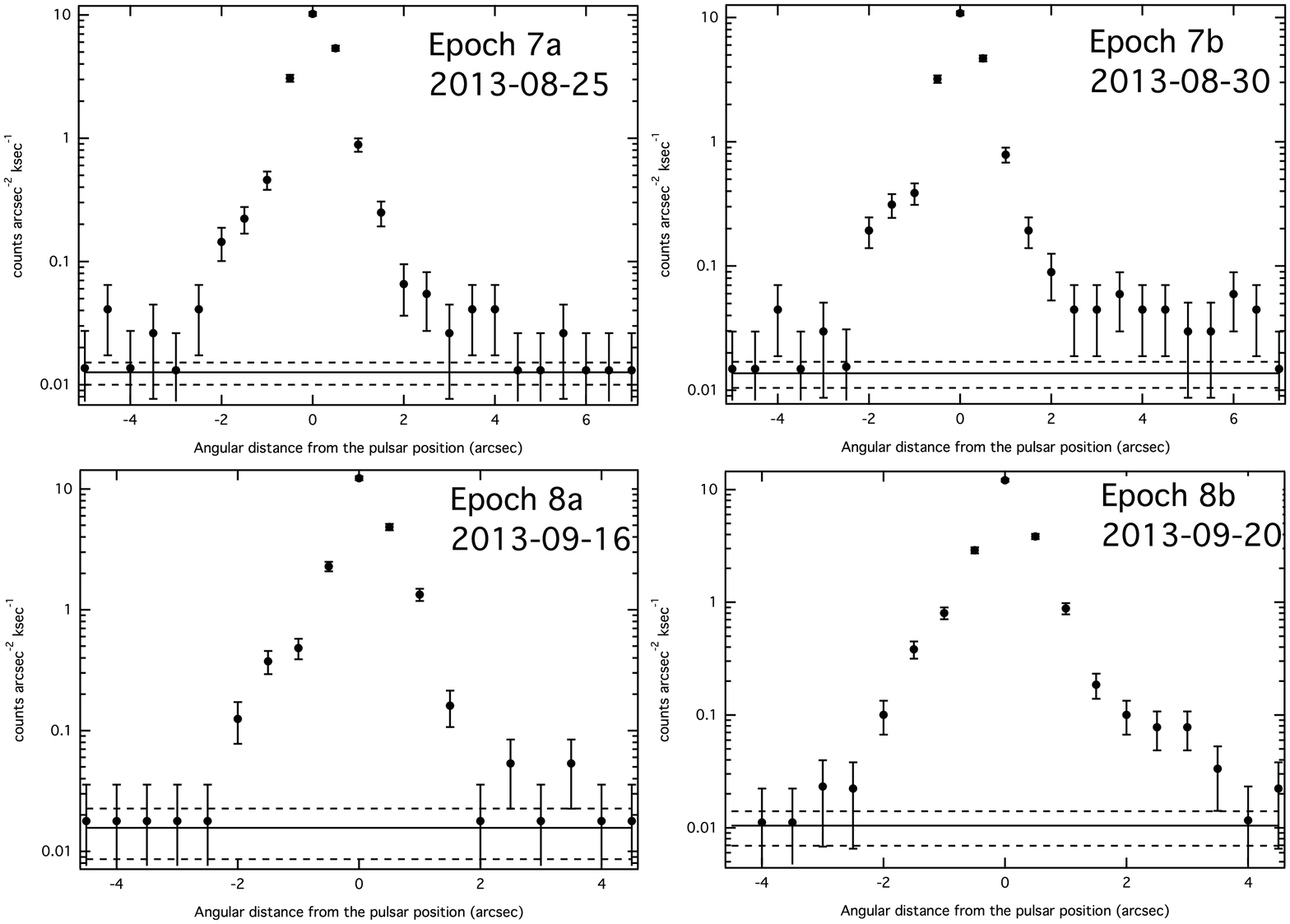}
\caption{The brightness profiles along the southeastern protrusion (cf. Fig.~\ref{subpix}) as computed from
the unsmoothed sub-pixel resolution images in Epochs 7a, 7b, 8a and 8b. The positive and negative direction of the $x$ axis correspond to the orientations of 
southeast and northwest from the pulsar position. The horizontal solid and dashed lines represent the background and its $1\sigma$ uncertainties at each epoch. 
The feature appears as the excess in the southeast with the length varies at different epochs.}
\label{jet_profile}
\end{figure*}

\subsection{Circumstellar Environment}
Inspecting Figure~\ref{axial}, we noted the presence of
extended emission of few arcsecond in the vicinity of the pulsar. For example, the protrusions in the directions that connected
to the outer tails in Epochs 7 and 8 are particularly prominent. In the following section, we will explore these small-scale features
in details.

We have performed a detailed morphological study of the circumstellar environment to search for the
possible diffuse X-ray feature around the pulsar. For each observation, we have produced a $30^{"}\times30^{"}$ exposure-corrected image
centered at the pulsar with sub-pixel resolution ($0.25^{"}\times0.25^{"}$ for one pixel).
We have identified the possible compact extended features around the pulsar in Epochs 4, 6, 7 and 8 through visual inspections of these images.
These extended feature candidates are shown in Figure~\ref{jet_candidate}.
All these features resemble a collimated protrusion extends to southeast direction that connected to the southern outer tail.

Posselt et al. (2017) have also spotted this feature in Epochs 4, 6 and 8 (cf. Sec. 2.3 in their paper). However, the detection significance has 
not been addressed. We proceeded to examine this putative feature with a more detailed analysis. Estimated with the nearby source-free background regions, 
we found the signal-to-noise ratios of the features
identified in Epochs 4 and 6 are only at the level of $\sim4.5\sigma$. According to our pre-defined detection criterion,
we are not able to firmly distinguish if they are
genuine or results of background fluctuations. For Epoch 7, we have investigated Epoch 7a and Epoch 7b separately.
Their images are shown in the upper panels of Figure~\ref{subpix}. A southeastern protrusion is noted in both Epoch 7a and Epoch 7b. 
The feature can be detected at a signal-to-noise ratio of $5.1\sigma$ in both epochs. 
We have also examined the robustness of these detections with various source-free backgrounds in nearby regions. 
We found the signal-to-noise ratios of the feature remain above our predefined threshold in both epochs.

For quantifying the length of the southeastern protrusion, we computed the brightness profiles from the unsmoothed sub-pixel resolution 
images in Epoch 7a and 7b which are shown in the upper panels of Figure~\ref{jet_profile}. The length of the feature is estimated by its profile extends in the 
southeastern direction from the pulsar before it falls into the background. 
Comparing Epoch 7a and Epoch 7b, the feature is found to be lengthened by $\sim2.5"$ within $\sim5$~days.
At a distance of 250~pc, the change of its physical length is $\sim9.4\times10^{15}$~cm.
This implies the variability of this feature occurred at a speed of $\sim0.8c$.
With the column absorption fixed at $1.1\times10^{20}$~cm$^{-2}$,
its spectrum can be described by a power-law model with a photon index of $\Gamma=1.16\pm0.48$. Its absorption-corrected flux is at
the level of $f_{x}\sim1.0\times10^{-14}$~erg~cm$^{-2}$~s$^{-1}$ (0.3-8~keV).

For Epoch 8, we have also separately analyzed the observations in 2013 September 16 (hereafter Epoch 8a) and 2013 September 20
(hereafter Epoch 8b). The close-up images of the pulsar obtained by these two observations are shown in the lower panels of Figure~\ref{subpix}.
While the protrusion is undetected in Epoch 8a,  it is found at a significance of $5.3\sigma$ in Epoch 8b. 
Compare the brightness profiles in Epoch 8a and 8b (lower panels in Figure~\ref{jet_profile}), it shows the protrusion was extended by $\sim2"$ in Epoch 8b.
These show the rapid variability of this feature at a timescale of few days as has been seen in Epoch 7.
Examining with various backgrounds sampled from different source-free regions, we found this result is robust.
With the column absorption fixed at $1.1\times10^{20}$~cm$^{-2}$,
its spectrum can be described by a power-law model with a photon index of $\Gamma=2.19^{+0.47}_{-0.43}$.
Its absorption-corrected flux is at the level of $f_{x}\sim6.8\times10^{-15}$~erg~cm$^{-2}$~s$^{-1}$ (0.3-8~keV).

In Figure~\ref{subpix}, we also noted that a counter-protrusion apparently extends toward northwestern direction in both epochs. Its brightness
is about a factor of two lower than that of the southeastern protrusion. However, such feature is
only significant at a $\sim2\sigma$ level and therefore we cannot confirm its existence.

\begin{figure}
\plotone{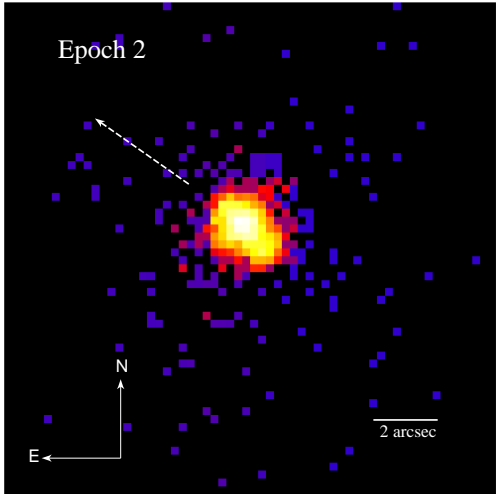}
\caption{Exposure-corrected sub-pixel resolution raw image (image bin-size=0.25$^{"}$)
in the energy range of 0.5-7~keV of a $15^{"}\times15^{"}$ field centered on
Geminga as observed by \emph{Chandra} ACIS-I on 27 August 2007 (i.e. Epoch 2). The distribution of the X-ray photons from the pulsar
is elongated along the orientation of its proper motion.}
\label{subpix_raw}
\end{figure}

We have also investigated the spatial nature of the X-ray emission at the pulsar position through a two-dimensional image fitting with the aid of \emph{Sherpa}.
We aim to probe if there is any indication for the compact nebula associated with Geminga. For each epoch,
we fit the sub-pixel resolution image with a composite model of a 2-D Gaussian model plus a horizontal plane to
account for the compact source and the background respectively. Except for Epoch 2, the spatial distributions of the X-ray photons from Geminga
in all the other observations conform with that expected from a point source. In Figure~\ref{subpix_raw}, the close-up unsmoothed image of Epoch 2 with a field
of $15^{"}\times15^{"}$ centered at the pulsar is shown. The distribution of the photons is apparently elongated.
The 2-D fitting yields a FWHM of $1.35^{"}\pm0.02^{"}$ with respect to the best-fit major axis
which is along the proper-motion direction.
We have subsequently examined the spectrum of the pulsar from Epoch 2 to see if there is any evidence of an additional contribution from the
compact nebula. Adopting the same spectral models used in Mori et al. (2014), we found that all the parameters inferred from fitting
the Epoch 2 spectrum are consistent with them and there is no indication that any extra spectral component is required.

We have also re-examined the arc-like feature in Epochs 1 and 2 as reported by Pavlov et al. (2010), we have adopted a similar source region in
their study (cf. Fig. 2 in their paper). We found the signal-to-noise ratios of this feature are only $\sim3.4-4\sigma$.
These are consistent with the results of Pavlov et al. (2010). However, it is below
the $5\sigma$ criterion we adopted to firmly discriminate genuine feature from the background fluctuation in this work. Also, we do not find
any solid evidence for such feature in the other observations. Therefore, we will not further consider this feature in our work.

\subsection{Outer Tails}
\begin{figure}
\plotone{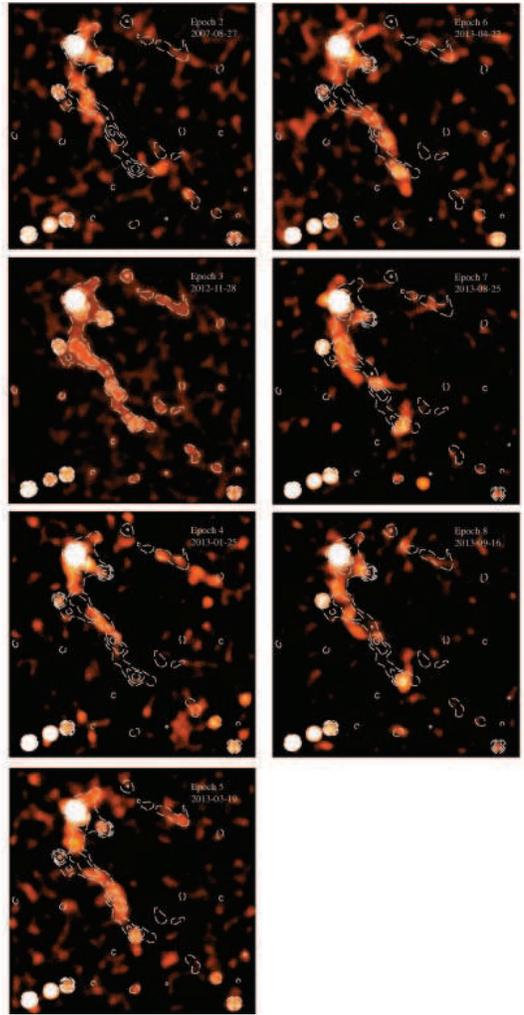}
\caption{A time sequence of \emph{Chandra} $5^{'}\times5^{'}$ exposure-corrected images (0.5-7~keV)
that cover the outer tails of Geminga. The observation in 2004 is excluded in this analysis as it did not cover the
whole feature. All images are smoothed wtih a Gaussian kernel of $\sigma=6^{"}$. The contours computed from the images obtained
in Epoch 3 are overlaid on all images for comparing the morphology of outer tails at different epochs.}
\label{outer_seq}
\end{figure}

For investigating the variability of both outer tails, Epoch 1 data is excluded in this analysis as it does not
cover the whole feature.
In Figure~\ref{outer_seq}, we have shown the time sequence of $5^{'}\times5^{'}$ exposure-corrected images from Epoch 2 to
Epoch 8. We computed the X-ray contours from the Epoch 3 image which has a high signal-to-noise ratio ($\sim9.4\sigma$) and the
whole feature is not intervened by any CCD gap. We adopted these contours as a reference and overlaid them on all images
to investigate if there is any morphological variations at different epochs.

In Epoch 4, both northern and southern tails are found to be apparently different from those in the other epochs.
The southern tail appears to be shorter in
Epoch 4. However, this is due to the fact that a large fraction of this feature fell in the CCD gap. On the other hand, the northern
tail appears to be longer in this epoch. But the signal-to-noise ratio of this feature in Epoch 4 is only $\sim2.3\sigma$ and
hence its variability is inconclusive. In view of this,
we conclude that the morphologies of both outer tails from Epochs 2 to 6 are found to be consistent.

\begin{figure*}
\plotone{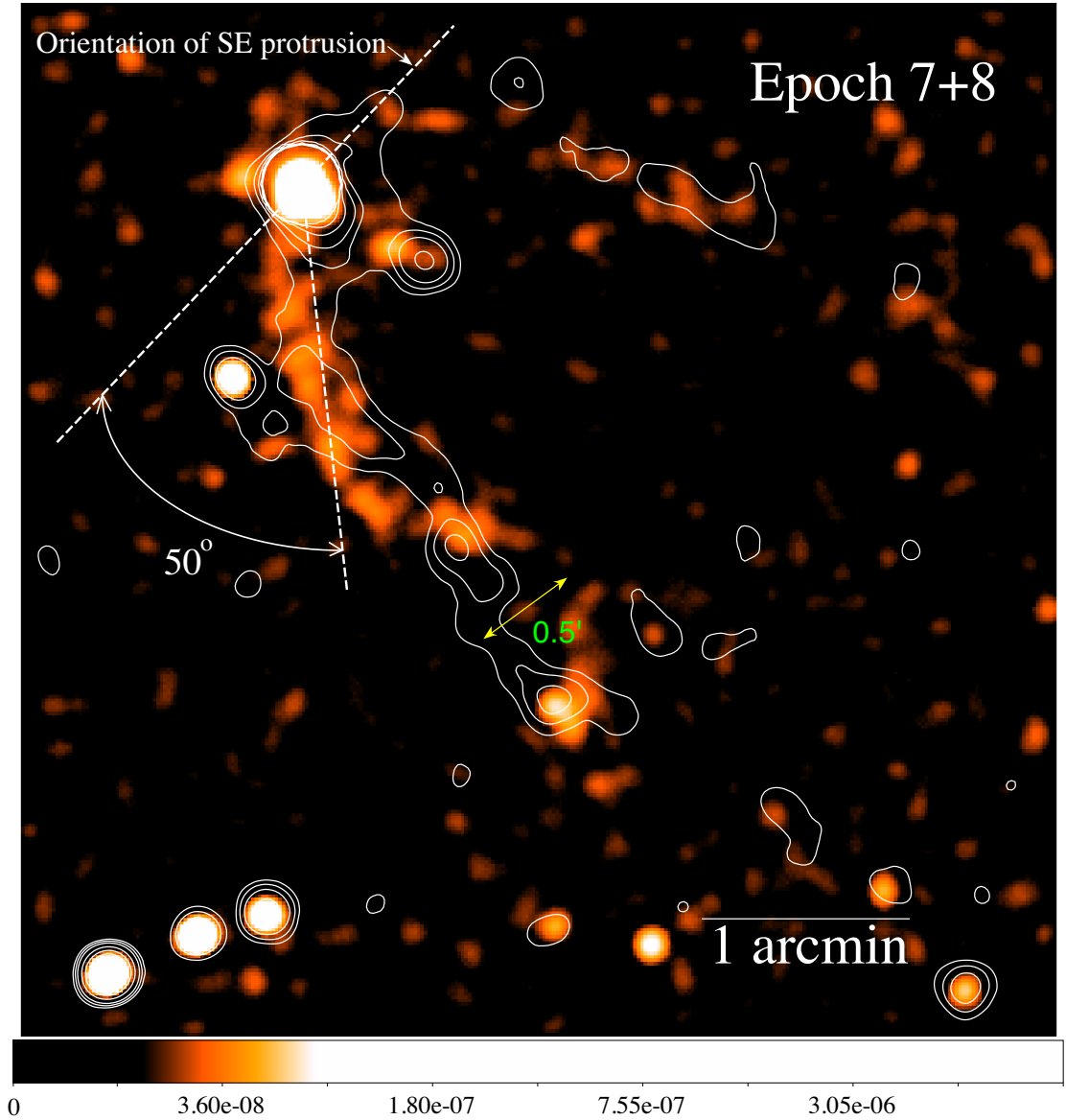}
\caption{$5^{'}\times5^{'}$ exposure-corrected image (0.5-7~keV; $1^{"}\times1^{"}$ for one pixel)
that cover the outer tails of Geminga as observed by {\it Chandra} in 2013 August and September (Epochs 7+8).
It has been smoothed wtih a Gaussian kernel of $\sigma=6^{"}$. The contours computed from the images obtained
in 2012 November/December (Epoch 3) are overlaid for visualizing the morphological variation of the outer tails.
The wiggling motion of the southern outer tail can be found by this comparison. The maximum lateral displacement 
of $\sim0.5^{'}$ at the rear part of the tail is shown by the yellow double-arrow. The deviation of the southern  
outer tail with respect to the southeastern protrusion is also illustrated.
The scale bar at the bottom shows the pixel values in the unit of photon cm$^{-2}$ s$^{-1}$.
See also Movie 4 (online supplementary material) for the animated comparison between the frames in Epoch 3 and Epochs 7+8.}
\label{outer_clear}
\end{figure*}

In Epochs 7 and 8, the southern tail is found to be significantly twisted with respect to the contours (see Fig.~\ref{outer_seq}).
A S-shaped structure can be clearly seen in these images.
Since such morphological change has been displayed in different images obtained in two consecutive epochs, indicating that it
is not a result of background fluctuations.
For highlighting the twisted feature, we combined the exposure-corrected images of Epochs 7 and 8 in Figure~\ref{outer_clear} so as to
enhance its significance. This morphological variation is different from that of axial tail,  
there is no propagation can be found in the southern outer tail as the far end of this feature appears to fix at the same position 
in Epoch 7+8 and Epoch 3. Instead, obvious lateral displacement can be seen by comparing these images (see also movie 4 in the 
supplementary material for the animated illustration).
Comparing the images in Epoch 3 and Epoch 7 (separated by $\sim9$ months), the maximum deviation of the southern tail is $\sim0.5^{'}$ 
(illustrated in Figure~\ref{outer_clear}). At a distance of 250~pc, this lateral displacement corresponds to a physical scale 
of $\sim10^{17}$~cm.
These imply that the twisting of the southern tail occurred at a speed of $\sim0.2c$.

To investigate if there is any short-term variability at the order of few days, which are found in the axial tail and
the circumstellar diffuse emission, we have also searched for the possible morphological variability of the outer tails among
the individual observations in Epochs 7 and 8. Nevertheless, we do not found any significant short-term variability.

For the spectral analysis, we firstly cross-checked the epoch-averaged outer tails obtained by Posselt et al. (2017) with the simultaneous 
fit of all the data from Epoch 2 to Epoch 8. 
For each epoch, we adopted a region so as to fit the morphology of each tails and avoid the CCD gaps.
$\Gamma=0.94\pm0.20$ (northern tail) and $\Gamma=1.02\pm0.09$ (southern tail) are derived 
from our independent analysis which are consistent with those obtained by Posselt et al. (2017) (cf. Tab.~3 in their paper). 
For the southern tail, 
while the flux is comparable with that obtained by Pavlov et al. (2010) based on Epoch 2 data,
the epoch-averaged photon index obtained by Posselt et al. (2017) and us is apparently harder than that reported by 
Pavlov et al. (2010), namely $\Gamma=1.59^{+0.30}_{-0.28}$. 
This indicates the possible spectral hardening in the later epochs. However, the search for the spectral variation of the 
outer tails has not been considered in Posselt et al. (2017).  

\begin{figure*}
\plotone{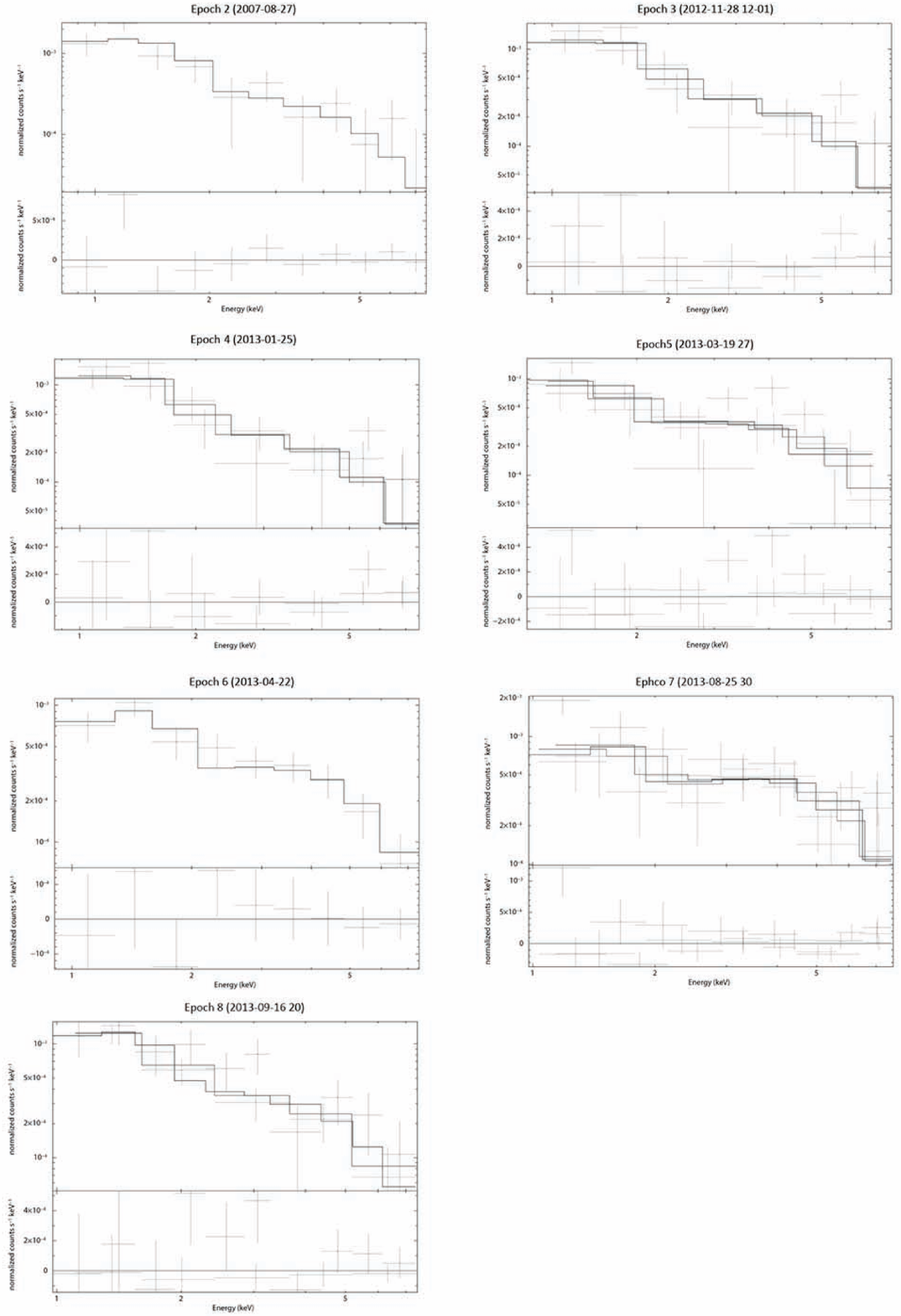}
\caption{Background-subtracted energy spectra of the southern outer tail ({\it upper panels}) with the best-fit power-law models (solid line) in different epochs.
The best-fit spectral parameters are given in Tab.~\ref{outer_spec}.
The fitting residuals are also shown ({\it lower panels}).}
\label{south_spec_plot}
\end{figure*}

We proceeded to examine the spectral properties of the southern tail in each epoch. The results are summarized in Table~\ref{outer_spec}. 
The corresponding background-subtracted spectra and the fitting residuals are shown in Figure~\ref{south_spec_plot}.
The spectral parameters obtained from Epoch 2 are consistent with those reported by Pavlov et al. (2010). 
In Epoch 4-7, the best-fit photon indices appear to become smaller. This explains the discrepancy between 
Posselt et al. (2017) and Pavlov et al. (2010).
Particularly in Epoch 7, its spectrum became the hardest among all epochs when the tail was significantly twisted (see Fig.~\ref{outer_clear}). 
Comparing with the epoch-averaged result, the photon index in this epoch is smaller by $\sim3\sigma$. 
In Epoch 8, the spectrum becomes softer again. 

For the northern tail,
its signal-to-noise ratio is $<4\sigma$ in images of all seven epochs.
And hence, we are not able to conclusively quantify if there is any morphological variability.

\begin{figure*}
\plotone{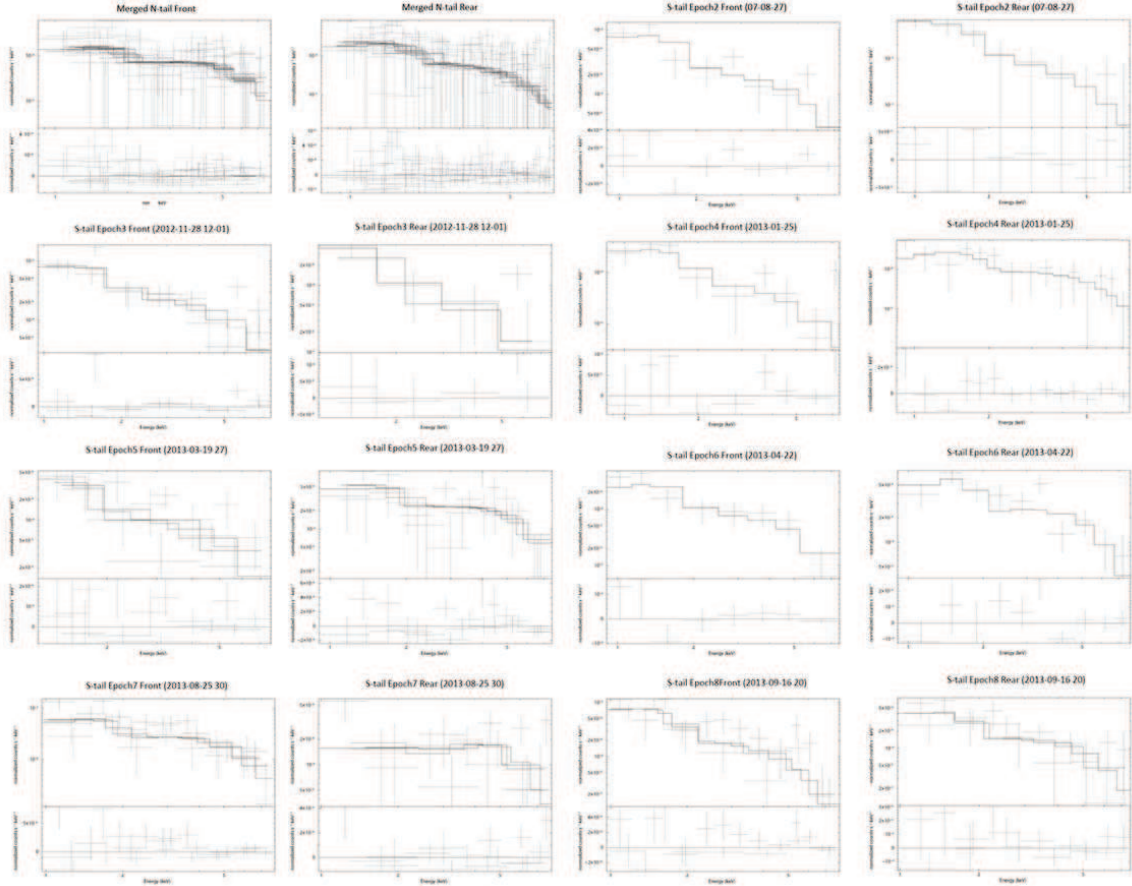}
\caption{Spatially-resolved background-subtracted energy spectra of northern and southern outer tails ({\it upper panels}) 
with the best-fit power-law models (solid line) in different epochs.
The best-fit spectral parameters are given in Tab.~\ref{outer_spec_resol}.
The fitting residuals are also shown ({\it lower panels}).}
\label{outer_spatial_plot}
\end{figure*}

We further investigated the spectral hardening of the southern tail. From Figure~\ref{feature}, it appears that its X-ray emission
is harder in the region further away from the pulsar than that close to the pulsar. 
This motivates us to performed a spatially-resolved analysis to investigate if there is any spatial variation of spectral properties and 
pinpoint the origin of the spectral hardening.
For each tail, we divided it into two halves with one closer to the pulsar (hereafter ``front") and the other further way from it (hearafter ``rear").
The results are summarized in Table~\ref{outer_spec_resol}.
The corresponding background-subtracted spectra and the fitting residuals are shown in Figure~\ref{outer_spatial_plot}.

For the southern tail, there is an indication of spectral softening along the southern tail found in Epochs 2 and 3. 
On the contrary, spectral hardening toward the end of the feature has been observed
from Epoch 4 to Epoch 8. Particularly in Epoch 7, the difference of $\Gamma$ inferred in the front and rear spectra is $\sim3\sigma$.

\begin{figure*}
\plotone{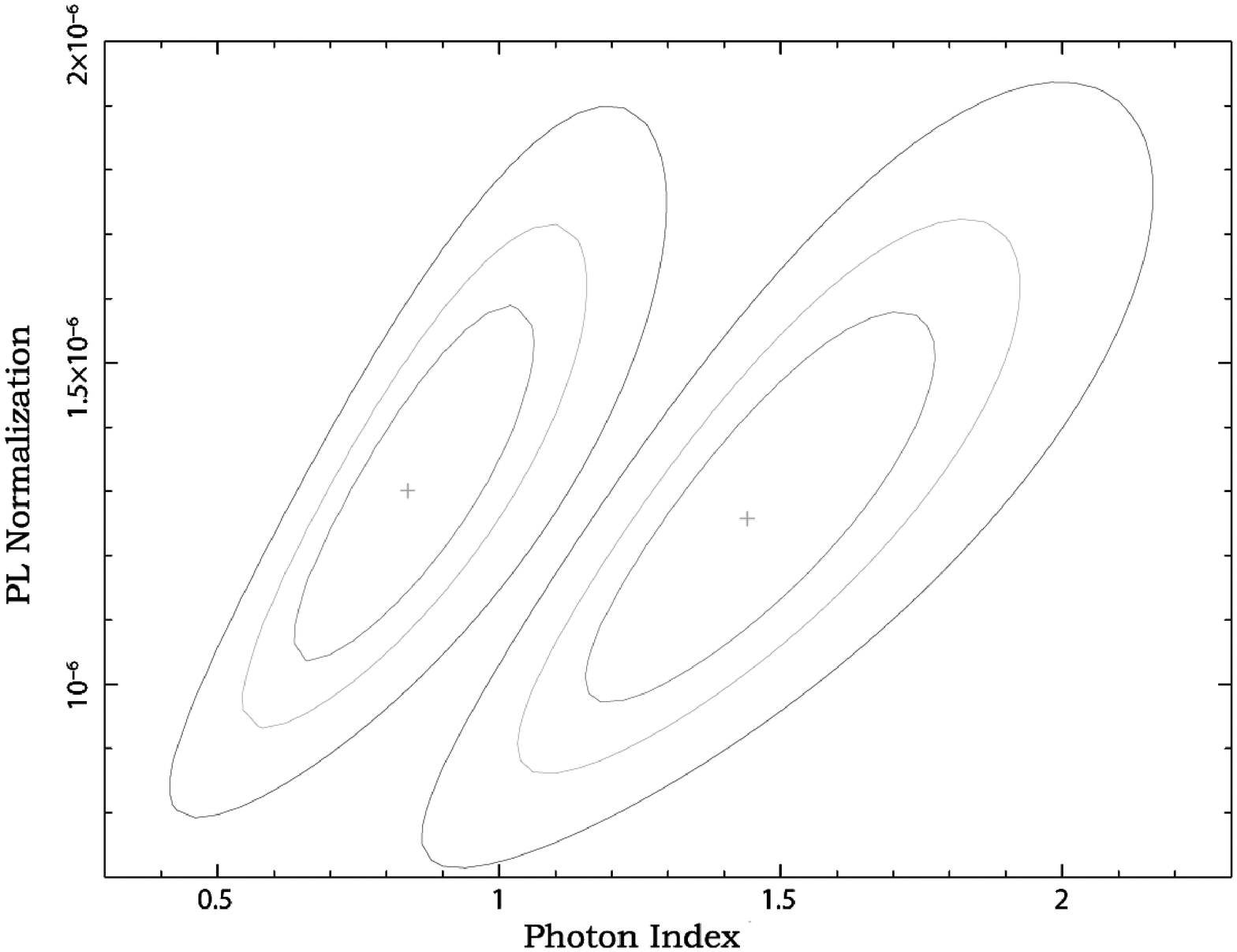}
\caption{$1\sigma$, $2\sigma$, $3\sigma$ 
confidence contours for the parameters of the ``front" part (right contours) and the ``rear" part (left contours) 
of the outer southern tail derived in the period of Epochs 4-8. Spectral hardening along the tail is noticed in this period.}
\label{south_contour}
\end{figure*}

We further examined the spectral hardening along the southern tail from Epoch 4 to Epoch 8. 
By simultaneously fitting their ``front" spectra in these five epochs, we obtained a photon index of $\Gamma=1.32^{+0.17}_{-0.16}$.
Similarly, it yields a photon index of $\Gamma=0.84\pm0.14$ for the ``rear" part. The corresponding confidence contours are shown 
in Figure~\ref{south_contour}, which clearly show the significant spectral variation along the southern outer tail. 

We have cross-checked all the aforementioned results of spectral hardening with backgrounds sampled from different low count regions. 
The results of all the independent spectral fits are found to be consistent within the tolerence of their statistical uncertainties.

We note that Posselt et al. (2017) did not find any spectral variation along the southern tail. This can be ascribed 
to the fact that they have only analyzed the combined spectra. As there are evidence of spectral softening in Epochs 2-3, 
the spectral variation can be averaged out in the combined spectra. 

We have also examined the front and rear spectra of the northern tail with all the observations combined, we have
noticed the best-fit photon index in the rear part $\Gamma\sim1.3$ is larger than that of the front part $\Gamma\sim0.6$.
This might indicate a spectral softening along the feature. However, in view of the larger statistical uncertainties,
we cannot unambiguously conclude this with the existing data.

\begin{table*}
\caption{Spectral properties of the southern tail in different epochs.}
\centering
\begin{tabular}{c c c c c c c c c}
\hline
\hline
 & $A_{S}$ & $N_{\rm net}$ & $N_{\rm bkg}$ & $\alpha$ & S/N & $\Gamma$ & $F_{\rm obs}$ & $L_{\rm unabsorb}$ \\
 & arcsec$^2$ & counts & counts &  & $\sigma$ & & $10^{-14}$ erg cm$^{-2}$ s$^{-1}$ & $10^{29}$ erg/s \\
\hline
Epoch 2 & 5881 & 198 & 1329 & 0.49 & 5.8 & $1.66^{+0.37}_{-0.32}$ & $3.12^{+0.70}_{-0.61}$ & $2.34^{+0.53}_{-0.45}$ \\

Epoch 3 & 5001 & 311 & 3172 & 0.20 & 9.4 & $1.43^{+0.27}_{-0.24}$ & $3.31^{+1.25}_{-0.50}$ & $2.48^{+0.93}_{-0.37}$ \\

Epoch 4 & 1816 & 126 & 3264 & 0.08 & 6.2 & $1.08^{+0.31}_{-0.29}$ & $1.86^{+0.39}_{-0.36}$ & $1.39^{+0.29}_{-0.27}$ \\

Epoch 5 & 3726 & 224 & 3079 & 0.14 & 8.3 & $0.87^{+0.24}_{-0.23}$ & $4.23^{+0.62}_{-0.59}$ & $3.16^{+0.46}_{-0.44}$ \\

Epoch 6 & 2486 & 279 & 4263 & 0.08 & 11.2 & $0.81\pm0.17$ & $4.17^{+0.52}_{-0.50}$ & $3.12^{+0.39}_{-0.38}$ \\

Epoch 7 & 4730 & 333 & 3887 & 0.14 & 10.7 & $0.41\pm0.20$ & $6.18^{+0.79}_{-0.76}$ & $4.62^{+0.59}_{-0.57}$ \\

Epoch 8 & 4485 & 291 & 5013 & 0.10 & 9.9 & $1.25^{+0.21}_{-0.20}$ & $3.90^{+0.51}_{-0.49}$ & $2.92^{+0.38}_{-0.36}$ \\
\hline
\hline
 \end{tabular}
\label{outer_spec}
 \end{table*}

\begin{table*}
\caption{Results from the spatially-resolved spectral analysis of the outer tails.}
\centering
\begin{tabular}{l c c c c c c c c}
\hline
\hline
 & $A_{S}$ & $N_{\rm net}$ & $N_{\rm bkg}$ & $\alpha$ & S/N & $\Gamma$ & $F_{\rm obs}$ & $L_{\rm unabsorb}$ \\
 & arcsec$^2$ & counts & counts &  & $\sigma$ &  & $10^{-14}$ erg cm$^{-2}$ s$^{-1}$ & $10^{29}$ erg/s \\
\hline
\multicolumn{7}{c}{Northern Tail}  \\
\hline
Merged (Front) & 1844 & 333 & 8972 & 0.18 & 7.1 & $0.61\pm0.27$ & $0.93^{+0.16}_{-0.15}$ & $0.69^{+0.12}_{-0.11}$ \\
Merged (Rear) & 3236 & 231 & 8972 & 0.31 & 3.7 & $1.30^{+0.34}_{-0.30}$ & $0.68^{+0.16}_{-0.15}$ & $0.51^{+0.12}_{-0.11}$ \\
\hline
\multicolumn{7}{c}{Southern Tail}  \\
\hline
Epoch 2 (Front) & 1974 & 123 & 1329 & 0.17 & 6.3 & $1.36^{+0.40}_{-0.34}$ & $2.10^{+0.54}_{-0.46}$ & $1.57^{+0.40}_{-0.34}$ \\
Epoch 2 (Rear) & 3282 & 64 & 1329 & 0.27 & 2.7 & $2.12^{+0.89}_{-0.62}$ & $1.01^{+0.38}_{-0.27}$ & $0.76^{+0.28}_{-0.20}$ \\
\\
Epoch 3 (Front) & 2951 & 239 & 3172 & 0.12 & 9.2 & $1.33^{+0.28}_{-0.25}$ & $2.34^{+0.41}_{-0.38}$ & $1.75^{+0.30}_{-0.28}$ \\
Epoch 3 (Rear) & 2149 & 102 & 2535 & 0.10 & 5.2 & $1.87^{+1.37}_{-0.81}$ & $0.71^{+0.31}_{-0.22}$ & $0.53^{+0.23}_{-0.16}$ \\
\\
Epoch 4 (Front) & 591 & 68 & 3264 & 0.03 & 5.4 & $1.73^{+0.52}_{-0.44}$ & $0.60^{+0.17}_{-0.14}$ & $0.45^{+0.12}_{-0.11}$ \\
Epoch 4 (Rear) & 1225 & 58 & 3264 & 0.05 & 3.7 & $1.07^{+0.45}_{-0.39}$ & $1.03^{+0.33}_{-0.29}$ & $0.77^{+0.24}_{-0.22}$ \\
\\
Epoch 5 (Front) & 658 & 71 & 3079 & 0.03 & 5.8 & $1.40^{+0.51}_{-0.46}$ & $1.08^{+0.24}_{-0.21}$ & $0.81^{+0.18}_{-0.16}$ \\
Epoch 5 (Rear) & 3068 & 153 & 3079 & 0.12 & 6.5 & $0.66\pm0.27$ & $3.00^{+0.57}_{-0.54}$ & $0.24^{+0.43}_{-0.40}$ \\
\\
Epoch 6 (Front) & 466 & 74 & 4263 & 0.01 & 6.4 & $1.28^{+0.44}_{-0.36}$ & $0.83^{+0.21}_{-0.19}$ & $0.62^{+0.16}_{-0.14}$ \\
Epoch 6 (Rear) & 2020 & 205 & 4263 & 0.06 & 9.3 & $0.70\pm0.20$ & $3.08^{+0.46}_{-0.44}$ & $2.31^{+0.35}_{-0.33}$ \\
\\
Epoch 7 (Front) & 2476 & 203 & 3887 & 0.08 & 8.9 & $0.70^{+0.20}_{-0.21}$ & $3.29^{+0.48}_{-0.47}$ & $2.46^{+0.36}_{-0.35}$ \\
Epoch 7 (Rear) & 2678 & 129 & 3887 & 0.08 & 6.0 & $-0.13^{+0.50}_{-0.59}$ & $2.66^{+0.68}_{-0.62}$ & $1.99^{+0.51}_{-0.475}$ \\
\\
Epoch 8 (Front) & 2298 & 173 & 5013 & 0.05 & 8.1 & $1.57^{+0.32}_{-0.29}$ & $1.84^{+0.32}_{-0.29}$ & $1.37^{+0.24}_{-0.22}$ \\
Epoch 8 (Rear) & 2189 & 118 & 5013 & 0.05 & 6.0 & $1.01^{+0.31}_{-0.29}$ & $1.67^{+0.37}_{-0.34}$ & $1.25^{+0.27}_{-0.26}$ \\
\hline
\hline
 \end{tabular}
\label{outer_spec_resol}
 \end{table*}

\section{SUMMARY \& DISCUSSIONS}
In this investigation, we have searched for the X-ray variabilities of the PWN powered by Geminga. Apart from cross-checking the 
epoch-averaged results recently reported by Posselt et al. (2017), we report a number of new findings in our variability analysis, 
which are summarized as follows:

\begin{enumerate}
\item
Discovery of spectral and morphological variations of axial tail in 2013 August/September, which took place 
at a timescale as short as a few days (Fig.~\ref{contour} \& \ref{subpix}). 
\item
Confirming the existence of the variable southeastern protrusion that connects the pulsar and the southern outer tail 
(Fig.~\ref{subpix} \& \ref{jet_profile}).
\item
The spatial distribution of the photons from the pulsar is found to be elongated in 2007 August (Fig.~\ref{subpix_raw}).
\item
Confirming the wiggling of the southern outer tail that took place at a timescale of few months (Fig.~\ref{outer_clear}).
\item 
Discovery of the spectral hardening of the southern outer tail from 2013 January to 2013 August (Tab.~\ref{outer_spec}).
We further identified the spectral hardening occured at the rear end of the feature (Fig.~\ref{feature} \& \ref{south_contour}). 
\end{enumerate}

All these results can be interpreted in the context of a general model of PWN structure which consists of a torus around the pulsar, 
jets along the spin axis and a synchrotron nebula resulted from the post-shock flow (cf. Gaensler \& Slane 2006 for a detailed review). 
In the following discussion, we assume the compact elongated structure (Fig.~\ref{subpix_raw}), southeastern protrusion+outer tails 
(Fig.~\ref{subpix} \& Fig.~\ref{outer_clear}) and the axial tail (Fig.~\ref{axial} \& Fig.~\ref{subpix}) correspond to the torus, 
jets and the bow shock nebula respectively.

Among all these, the most spectacular results are the morphological and spectral variability of the southern outer tail. 
Both outer tails can be interpreted as bipolar outflows from the pulsar that are bent by
the ISM ram pressure. Such significant jet-bending has not been seen in other PWNe.
Our interpretation is supported by the discovery of the collimated southeastern protrusion which is connected to the
southern outer tail (Fig.~\ref{subpix}). The protrusion is likely to be a collimated jet powered by the pulsar.

In this scenario, the significant bending of the jets indicates the ram pressure of the ISM is comparable with that of the jet.
This structure is similar to the bent jets observed in some radio galaxies (e.g. McBride \& McCourt 2014).

The ram pressure of the jet $P_{j}$ can be expressed as:
\begin{equation}
P_{j}\sim\frac{\dot{E}}{\pi\theta_{j}^{2}r_{j}^{2}c}
\end{equation}

\noindent where $\theta_{j}$ and $r_{j}$ are the open angle and the length of the jet respectively. On the other hand, the ram pressure of the
ISM $P_{a}$ is given by:\begin{equation}
P_{a}\sim\rho_{\rm ISM}v_{\rm psr}^{2}
\end{equation}

\noindent where $\rho_{\rm ISM}$ and $v_{\rm psr}$ are the ISM density and the pulsar's velocity respectively. For $\rho_{\rm ISM}\sim2\times10^{-24}$~g~cm$^{-3}$ and
$v_{\rm psr}\sim200$~km~s$^{-1}$, the ISM ram pressure is at the order of $P_{a}\sim10^{-9}$~erg~cm$^{-3}$.
Assuming $P_{j}\sim P_{a}$ and $r_{j}\sim0.1$~pc, it implies the open angle of the jet with a
few degrees.

From the theory of jet bending model of galaxies (Begelman et al. 1979), the curvature radius $R_{c}$ of the bent jet can be expressed as:

\begin{equation}
R_{c}\sim\frac{P_{j}}{P_{a}}h
\end{equation}

\noindent where $h$ is the scale height of the jet perpendicular to the motion of the jet. For $P_{j}\sim P_{a}$, 
$R_{c}/h\sim1$ implies a bending angle of $\sim60^{\circ}$
which is not far away from the observed deviation (see Figure~\ref{outer_clear}).

In Epochs 7a, 7b and 8b, we also noted that a counter-protrusion apparently extends toward northwestern direction.
Its brightness is about a factor of two lower than that of the southeastern protrusion. The difference of brightness between this putative counterjet
and the southeastern protrusion can be explained by the Doppler boosting. We can place a constraint on the angle $\zeta$
between the spin axis and the line of sight. The brightness ratio $f_{b}$ can be written as:

\begin{equation}
f_{b}=\left(\frac{1+\beta\cos\zeta}{1-\beta\cos\zeta}\right)^{\Gamma +2}
\end{equation}

\noindent where $\Gamma$ is the photon index of the jet and $\beta$ is the outflow velocity in the unit of $c$.

From comparing the images in Epochs 7a/7b and Epochs 8a/8b, $\beta$ is found to be $\sim0.7-0.8$. In view of the low significance
of the counterjet ($\sim2\sigma$),
we consider a lower limit of $f_{b}$ to be $>2$. The photon index of the jet has a large uncertainties and it might
vary at different epoch. It spans a range of $\Gamma\sim0.7-2.7$ in Epoch 7b and 8b. With these observational inputs, we place an upper limit
of $\zeta$ to be $<80^{\circ}$. This is close to the value inferred by modeling the $\gamma-$ray light curve of Geminga (Zhang \& Cheng 2001).

The orientation of the jet and counterjet is almost perpendicular to the proper motion direction.
Assuming the spin axis of Geminga is aligned with jet/counterjet, its spin vector and the velocity vector are apparently
orthogonal to each other. This is different from the spin-velocity alignment as seen in many other pulsars (Noutsos et al. 2012).
The orthogonal spin-velocity configuration of Geminga might be a result of a single localized momentum impulse imparted
on the neutron star when it was born (Spruit \& Phinney 1998).

Interpreting this protrusion as a jet suggests the possible presence of a torus around the pulsar. A torus plus jet structure has been
seen in many other pulsars such as Crab (Weisskopf et al. 2000; Gaensler \& Slane 2006).
In searching for such compact structure, we have identified an elliptical feature
around the pulsar in Epoch 2 (Fig.~\ref{subpix_raw}).

This feature can be interpreted as a tilted torus around Geminga. This scenario is supported by the presence of the
jet-like protrusion along the toroidal axis. 
Relativistic magnetohydrodynamic (RMHD) simulations have shown that the energy of the pulsar wind is concentrated toward 
the its equator. When the wind interacts with the circumstellar medium, a termination shock can emerge. In the downstream 
of this shock, plasma flows into the equatorial region and thus forms a torus (cf. Fig.~4 in Porth et al. 2017). 
The plasma also flows to the polar region. In the presence of the magnetic hoop stress, the jets perpendicular to the torus 
can be formed (for a recent review on the RMHD study of PWN, please refer to Porth et al. 2017). 

Under this assumption, the termination shock radius $R_{s}$ can be estimated by the balance
between the ram pressure of the pulsar wind and the magnetic pressure $B^{2}/8\pi$ of the PWN:

\begin{equation}
R_{s}\simeq\left(\frac{\dot{E}}{B^{2}c}\right)^{1/2}.
\end{equation}

\noindent At a distance of 250 pc, three Gaussian sigma, $1.5^{"}$, inferred from image fitting
corresponds to a physical size of $5.6\times10^{15}$~cm which is about two orders smaller than the X-ray ring of Crab
Nebula. Taking this as the estimate of $R_{s}$, it implies a magnetic field of $B\sim195\mu$G which is typical for PWNe (Gaensler \& Slane 2006).
As this elliptical feature can only be seen in one observation, it is very likely that the feature is also variable.
The torus of Crab has been found to shrink (Mori et al. 2004). In case that the elliptical feature of Geminga detected in 2007
August shrinks by at least a factor of 2, it will not be able to resolved by any state-of-the-art X-ray telescope.

Variability timescales ranging from a few days to a few months have been found in various spatial components of Geminga's PWN. 
These timescales should shed light and also put constraints on modelings of this complex system and magneto-hydrodynamical processes there.

The southeastern jet-like protrusion, for instance, is clearly variable. As it can only be detected in 2013 August and September, 
this indicates that the outflow was enhanced in these two epochs. This might be related to the observed wiggling motion 
of the outer southern tail in these epochs. 
A similar wiggling motion has also been observed in the jets of Vela and Crab (Pavlov et al. 2003; Mori et al. 2004). 
These variations are found at the order of few tens per cent of the speed of light.

The nature and origin of the variabilities of outer and axial tails were conjectured before. Pavlov et al. (2010) and Posselt et al. (2017), 
for instance, examined both the shell model of shocked wind and the jet-like outflow model; in the shell model the variabilities could result 
from shear instabilities, while in the outflow model they could be induced by internal instabilities or external environmental effects 
(that is, sweepback by external ram pressure). 

For the southern outer tail, different from the axial tail, it appears that the twisted structure does not have any propagation 
(see Fig.~\ref{outer_clear}). This indicates the nature of the variabilities of these two PWN components are different. The absence 
of propagation in the case of southern outer tail suggests it can possible be a helical structure resulted from torsional Alf\'{v}en wave. 
Similar phenomena have been seen in the solar coronal jets (Srivastava et al. 2017; Szente et al. 2017), though the scale of the Geminga jet 
is much larger.

While a rough scaling of $t\sim r/v$ ($r$ is the size of the spatial variabilities and $v$ is the corresponding propagation speed) 
seems to be applied, the ultimate understandings of the physical processes involved in the variabilities could possibly be achieved only 
through detailed magneto-hydrodynamical modelings including simulations that are beyond the scope of this paper.

\acknowledgments
CYH is supported by the National Research Foundation of Korea through grants 2014R1A1A2058590 and 2016R1A5A1013277.
JL is supported by BK21 plus program and 2016R1A5A1013277.
AKHK is supported by the Ministry of Science and Technology of Taiwan
through grants 103-2628-M-007-003-MY3 and 105-2112-M-007-033-MY2.
PHT is supported by the One Hundred Talents Program of the Sun Yat-Sen University.
JT is supported by the NSFC grants of China under 11573010
KSC is supported by a 2014 GRF grant of Hong Kong Government under HKU 17300814P.
DR is supported by the National Research Foundation of Korea through grants 2016R1A5A1013277.


\end{document}